\documentclass[,superscriptaddress,single-spaced,floatfix,linenumbers,twocolumn]{revtex4}
\usepackage{graphicx,amsmath,bbm,bm,array,subfigure}
\usepackage{appendix}
\usepackage{lipsum}
\usepackage{dsfont}
\usepackage{calrsfs}
\usepackage[linktocpage=true,colorlinks=true,linkcolor=blue,citecolor=blue,urlcolor=blue]{hyperref}
\usepackage{MnSymbol}
\def\nn{\nonumber\\}

\begin{document}
%\maketitle
\title {The fate of nonlinear perturbations near the QCD critical point}
\author{Golam Sarwar}
\email{golamsarwar1990@gmail.com }
\affiliation{Department of Physics,University of Calcutta, 92, A.P.C. Road, Kolkata-700009, India}
\author{Md Hasanujjaman}
\email{jaman.mdh@gmail.com}
\affiliation{Department of Physics, Darjeeling Government College, Darjeeling- 734101, India}
%%%%
\author{Mahfuzur Rahaman}
\email{mahfuzurrahaman01@gmail.com }
\affiliation{Variable Energy Cyclotron Centre, 1/AF Bidhan Nagar, Kolkata- 700064, India}
\affiliation{Homi Bhabha National Institute, Training School Complex, Mumbai - 400085, India}
 \author{Abhijit Bhattacharyya}
\email{abhattacharyyacu@gmail.com}
\affiliation{Department of Physics,University of Calcutta, 92, A.P.C. Road, Kolkata-700009, India}
\author{Jan-e Alam}
\email{jane@vecc.gov.in}
\affiliation{Variable Energy Cyclotron Centre, 1/AF Bidhan Nagar, Kolkata- 700064, India}
\affiliation{Homi Bhabha National Institute, Training School Complex, Mumbai - 400085, India}
%\maketitle
%\date{\today} 
\def\zbf#1{{\bf {#1}}}
\def\bfm#1{\mbox{\boldmath $#1$}}
\def\hf{\frac{1}{2}}
\def\sl{\hspace{-0.15cm}/}
\def\omit#1{_{\!\rlap{$\scriptscriptstyle \backslash$}
{\scriptscriptstyle #1}}}
\def\vec#1{\mathchoice
{\mbox{\boldmath $#1$}}
{\mbox{\boldmath $#1$}}
{\mbox{\boldmath $\scriptstyle #1$}}
{\mbox{\boldmath $\scriptscriptstyle #1$}}
}
\def \beq{\begin{equation}}
\def \eeq{\end{equation}}
\def \beqa{\begin{eqnarray}}
\def \eeqa{\end{eqnarray}}
\def \pd{\partial}
\def \nn{\nonumber}
\begin{abstract}
The impact of the QCD critical point on the propagation of nonlinear waves
has been studied. The effects have been investigated within the scope of second-order causal dissipative hydrodynamics by incorporating the critical point into the equation of state, and the scaling behaviour of transport coefficients and of thermodynamic response functions. Near the critical point, the nonlinear waves are found to be significantly damped which may result in the disappearance of the Mach cone effects of the away side jet. Such damping may lead to enhancement in
the fluctuations of elliptic and higher flow coefficients. Therefore, the disappearance of Mach cone effects and the enhancement of fluctuations in flow harmonics in the event-by-event analysis may be considered as signals of the critical endpoint.
\end{abstract}

%\pacs{12.38.Mh, 12.39.-x, 11.30.Rd, 11.30.Er}
\maketitle

%\linenumbers
{\it \textbf{I. Introduction}: }
Relativistic Heavy Ion Collider Experiment (RHIC-E) is an excellent tool to explore the 
rich phase structure of Quantum Chromodynamics (QCD) under extreme conditions of 
temperature ($T$) and baryonic chemical potential ($\mu$). One of the main features
of the QCD phase diagram is the presence of the Critical Endpoint (CEP), 
where the first order phase transition from quark matter to hadronic matter  
terminates~\cite{Forcrand,Aoki, Endrodi,PNJL10,PQM1} and the transition becomes a 
crossover~\cite{Fodor,AsakawaYazaki,Halasz}. The search for the {\it elusive} CEP 
is one of the active areas of contemporary research~\cite{Caines}. 
Its exact location in the QCD phase diagram is not known
from the first principle (lattice QCD based) calculation due to the well known
sign problem for spin half particles. The theoretical 
prediction of the position of the CEP based
on effective models is still ambiguous as its location depends on the 
parameters of the models used.
On the experimental front, the beam energy scan (BES) program is being run with 
fine-tuning of the centre of mass energy $(\sqrt{s})$, such that if the quark-gluon plasma
(QGP) passes through the CEP, the effects should be reflected on the 
particle spectra. However, for experimental detection of the
CEP, theoretical investigations are required to understand its imprints on the data.

In this regard, mainly fluctuations of various thermodynamic quantities 
are studied~\cite{NahrangBluhm}, as the CEP is characterised by large 
fluctuations~\cite{Stanley} 
resulting from the diverging nature of the correlation 
length ($\xi$)~\cite{Kardar,Shankar} (see~\cite{Berdnikov,Stephanov2}
for RHIC-E related studies). 
However, apart from the studies of 
the fluctuations, the prediction on 
the fate of perturbations and their consequences on various 
experimental observables are 
significantly important for the detection  
of CEP. In RHIC-E, partons (quarks and gluons) 
are produced with a wide range of transverse momentum ($p_T$). 
Partons with relatively lower $p_T$,
on subsequent scattering, produce a locally thermalized 
hot medium of QGP, 
whereas, the high $p_T$ partons do not  
contribute in the medium formation, but they pass through the medium as 
jets with associated radiated partons. While propagating, 
these jets produce disturbances in the medium through interaction. 
The partons, especially the supersonic ones, can produce perturbations 
in the medium leading to nonlinear waves. Apart from the disturbances 
generated by the jets, quantum fluctuations 
leading to inhomogeneity in the medium may serve as perturbations 
in the hydrodynamically evolving medium. 
The hydrodynamic response of the medium to such perturbations, 
are reflected on the spectra of the produced hadrons at the freeze-out 
hypersurface and on other penetrating particles like photons and lepton pairs
emitted throughout the evolution history of the fireball. 
Specifically, the appearance of two maxima 
at $\Delta \phi=\pi \pm 1.2$ radian in the quenched away side jet or the double-hump 
in the correlation function of the jet, the structure is explained as the effect of the Mach 
cone produced due to hydrodynamic response to the perturbation created 
by jets~\cite{shuryaknlw}. The momentum anisotropy, which is quantified as flow 
harmonics of the produced particle is attributed to the hydrodynamic response of the QGP 
to the initial geometry. During the expansion and cooling, when 
the system makes a transition to the hadronic phase, these anisotropies 
get transmitted into hadronic momentum spectra through momentum conservation.

The damping of the Mach cone is connected to the hydrodynamic response. Any change in the nature of QGP medium and in the nature of the transition from QGP to hadron e.g. presence of the CEP 
will affect the response. In general, the hydrodynamic response can be treated as either linear or nonlinear depending on the magnitude of the perturbations. 
A small disturbance is treated as a linear 
perturbation~\cite{Shuryak1,Shuryak2,Rafiei:2016zxk,hasan1,hasan2,kunihiro} 
whereas a relatively large disturbance (of the order of unperturbed value) should be
treated as nonlinear perturbation~\cite{Raha1,Raha2, Fogaca1, Fogaca2}. The development of the momentum anisotropy can be accounted mostly as a linear response to the initial eccentricity with a small contribution from nonlinear effects~\cite{v2e2,v2ne2,v2ne22}. Moreover, the produced Mach front has been found to travel as a shock front~\cite{akc}. The propagation of this shock front is controlled by the hydrodynamic response which may be linear or nonlinear. Due to the large amplitude of this shock front, one expects the perturbation to be nonlinear in nature~\cite{Osbourne}. In Refs.~\cite{Betz1, Betz2, Li} the presence of Mach cone effect on the away side jet in two particles and three-particle correlations have been explained by using the non-dissipative property of the nonlinear waves.

The fate of nonlinear waves in QGP has been studied 
by Fogaca et al.~\cite{Fogaca1,Fogaca2} by
using Navier-Stokes (NS) and Israel-Stewart (IS)~\cite{Israel} 
hydrodynamics with the inclusion of shear viscous effects only. 
They found that despite the presence of shear viscous effects the nonlinear waves survive. The nonlinear 
perturbations induced by jets have been considered to be 
responsible for broadening of the away side jet 
~\cite{Fogaca1}. The effects of linear response have already been investigated and found to be suppressed near the CEP~\cite{kunihiro,hasan1,hasan2}. 
In this regard, however, it will be pertinent to investigate the fate of the nonlinear perturbations in presence of the CEP.
This is particularly important because physical processes like the deposition of energy by jets can be very large which may ignite nonlinear effects. In such cases,  
the nonlinear wave may play a dominant role in the formation of the Mach-cone. 
Therefore, the study of the effects of  
nonlinear perturbations in QGP under the influence 
of the CEP is crucial. 

The understanding of the broadening of 
the away side jet and the fate of conical flow or double-hump in the correlation function
will be useful for the critical point search. 
In fact, the fate of the nonlinear perturbations
due to the presence of CEP has not been addressed till date
to the best of our knowledge. 
In this work, we address the propagation of nonlinear waves in
the presence of the CEP.
To study the role of the CEP on the nonlinear perturbations,
the inclusion of all the relevant dissipative coefficients are
crucial as some of these are known to 
diverge near the CEP~\cite{Kharzeev,Karsch,Ryu,Kapusta,Martinez}. 
However, the equations governing the evolution of
the nonlinear perturbations in the presence of 
transport coefficients {\it i.e.} shear viscosity ($\eta$),
bulk viscosity ($\zeta$) and 
thermal conductivity ($\kappa$) are not readily available
within the scope of the second-order causal fluid dynamics
and hence, we deduce these equations in the present work. %The effects of bulk ($\zeta$) and shear ($\eta$) viscosities and the thermal conductivity ($\kappa$) are considered in the present work, in contrast to the previous work~\cite{Fogaca1} where the role of $\zeta$ and $\kappa$ were not taken into consideration. The roles of $\zeta$ and $\kappa$ might be crucial as these are expected to get enhanced near the CEP~\cite{Kharzeev,Karsch,Ryu,Kapusta}. \\
%%%

{\it \textbf{II. Non-linear wave equations}:} The nonlinear evolution equations for perturbation of hydrodynamic fields in the QGP can be derived from the relativistic viscous hydrodynamic equations. In NS hydrodynamics the 
dissipative flux is assumed to be proportional to the first-order gradient of the hydrodynamic fields which leads to acausal and unstable solutions. Though there are frame stabilized 
first-order relativistic hydrodynamics~\cite{Bemfica,Kovtun,Das}, it is not simultaneously casual and stable 
for a non-conformal system i.e., a system with quasi-particles with non-zero mass~\cite{arpan}. 
Therefore, for a consistent hydrodynamic description of QGP, 
second-order casual dissipative hydrodynamics of 
IS is considered as appropriate.

Here, we use the second-order theory {\it i.e} IS hydrodynamics, 
which respects causality and provides stable solutions. 
In general, there are two choices of frames of reference which are mostly used, 
namely Landau-Lifshitz (LL)~\cite{Landau} and Eckart frames~\cite{Eckart}.
The LL frame represents a local rest frame where the energy dissipation is zero but the net charge dissipation is non-zero. Whereas, 
the Eckart frame represents a local rest frame where the net charge 
dissipation is vanishing but the energy dissipation is non-vanishing. 
We adopt the Eckart frame of reference~\cite{Eckart} here. 
The energy-momentum tensor(EMT) and particle current can be written 
in the Eckart frame as:
\beqa
T^{\mu\nu}=\epsilon u^{\mu}u^{\nu}-p \Delta^{\mu\nu}+\Delta T^{\mu\nu}; \hspace{0.3cm}
J^{\mu}&=&nu^{\mu}
\eeqa
where, $\epsilon$, $p$ and $u^\mu$ are respectively the local energy density, 
pressure and fluid four velocity.
$\Delta T^{\mu\nu}$ is the dissipative part of the EMT.
We choose the Minkowski metric as: 
$g^{\mu \nu}=(1,-1,-1,-1)$ with 
$u^{\mu}u_{\mu}=1$ and 
$u^{\mu}\partial_{\nu}u_{\mu}=0$.
The projection operator normal to the four velocity is defined as 
$\Delta^{\mu\nu}=g^{\mu\nu}-u^{\mu}u^{\nu}$, such that $\Delta^{\mu\nu}u_{\nu}=0$ 
and $\Delta^{\mu\nu}\Delta_{\mu\nu}=3$. 
The symmetric traceless projection normal to the fluid four velocity is defined as
$\Delta^{\mu\nu}_{\alpha\beta}=\frac{1}{2}(\Delta^{\nu}_{\alpha}\Delta^{\mu}_{\beta}
+\Delta^{\mu}_{\alpha}\Delta^{\nu}_{\beta}-\frac{2}{3}\Delta^{\mu\nu}\Delta_{\alpha\beta})$.

The dissipative part of EMT can be written in terms of scalar, vector and tensor as:
\beqa
\Delta T^{\mu\nu}=-\Pi \Delta^{\mu\nu}+u^{\mu}q^{\nu}+u^{\nu}q^{\mu}+\pi^{\mu\nu}
\eeqa
 Therefore, the EMT in Eckart frame is:
 \beqa
 T^{\mu\nu}=\epsilon u^{\mu}u^{\nu}-(p+\Pi) \Delta^{\mu\nu}+u^{\mu}q^{\nu}+u^{\nu}q^{\mu}+\pi^{\mu\nu}
 \eeqa
The vector and tensor forms of dissipation are considered to be 
non-existent 
in the frame of fluid velocity, such that $u_{\mu}q^{\nu}=0$, $u_{\mu}\pi^{\mu\nu}=0$. 
The scalar dissipation ($\Pi$) corresponds to the non-equilibrium pressure perturbation which
can not be related to energy density through
the equation of state (EoS). The scalar dissipation is related to volume expansion of fluid that triggers small non-equilibrium perturbation which drives the system to a new state of equilibrium leading to dissipative correction. Energy density, pressure, and the dissipative fluxes 
are related to EMT by $u_{\mu}T^{\mu\nu}u_{\nu}=\epsilon, q_{\alpha}=u_{\mu}T^{\mu\nu}\Delta _{\nu\alpha}=u_{\mu}\Delta T^{\mu\nu}\Delta _{\nu\alpha}, p+\Pi=-\frac{1}{3}\Delta_{\mu\nu}T^{\mu\nu}, \Pi=-\frac{1}{3}\Delta_{\mu\nu}\Delta T^{\mu\nu}, u_{\mu}\Delta T^{\mu\nu}=q^{\nu}$.
The conservation of energy-momentum and the net charge (net baryon number here) 
density are given by the following equations:
\beqa
\partial_{\mu}T^{\mu\nu}=0; \hspace{0.3 cm} \partial_{\mu} J^{\mu}=0
\label{eq17}
\eeqa
The dissipative fluxes~\cite{Israel} are given by
\beqa
\Pi&=& -\frac{1}{3}\zeta[\pd_{\mu}u^{\mu}+\beta_{0}D\Pi-\tilde{\alpha_{0}}\pd_{\mu}q^{\mu}] \nn\\
%%%
\pi^{\lambda \mu}&=&-2\eta \Delta^{\lambda\mu\alpha\beta}\Big[\partial_{\alpha}u_{\beta}+\beta_{2}D\pi_{\alpha\beta}-\tilde{\alpha_{1}}\partial_{\alpha}q_{\beta}\Big]\nn\\
%%%
q^{\lambda}&=&-\kappa T\Delta^{\lambda\mu} [\frac{1}{T}\partial_\mu T +Du_{\mu}+\tilde{\beta_1} D{q_\mu}-\tilde{\alpha_0}\partial_\mu \Pi \nn\\
&&-\tilde{\alpha_1}\pd_{\nu}\pi ^{\nu}_{\mu} ] 
\label{eq11}
\eeqa
where $D\equiv u^\mu\partial_\mu$ is the co-moving derivative.
In the local rest frame (LRF) $D\Pi =\dot{\Pi }$, 
representing the time derivative. The relaxation times 
for the 
bulk pressure ($\tau_{\Pi}$), the heat flux ($\tau_q$) and the shear tensor ($\tau_{\pi}$) 
are given by~\cite{muronga} 
$\tau_{\Pi}=\zeta \beta_0, \,\,\,\,\tau_q=k_BT\beta_1$, and $\,\,\,\,\tau _{\pi}=2\eta \beta_2$. 
The relaxation lengths which couple the heat flux and bulk pressure 
($l_{\Pi q}, l_{q\Pi} $), the heat flux and shear tensor $(l_{q\pi}, l_{\pi q})$ 
are defined as:  
$l_{\Pi q}=\zeta \alpha_0,\,\,\,\, l_{q\Pi}=k_B T \alpha _0, \,\,\,\,l_{q\pi}=k_BT\alpha_1$, and $\,\,\,\, 
l_{\pi q}=2\eta \alpha_1 $.
%%%%%%%%%%
%%%%%%%%%%%%%%%%%%%%%%%%%%%%%%%%%%%%%%%%%%%%
The quantities, $\beta _0,\beta_1,\beta_2$ are called the relaxation coefficients, and $\alpha_0$, $\alpha_1$ are coupling coefficients. The quantities
$\tilde{\alpha_0}$, $\tilde{\alpha_1}, \tilde{\beta_{1}}$ 
in Eq.\eqref{eq11} 
are related to relaxation and coupling coefficients as 
$\tilde{\alpha_{0}}-\alpha_{0}=\tilde{\alpha_{1}}-{\alpha_{1}}=-(\tilde{\beta_{1}}-\beta_{0})
=-[(\epsilon+p)]^{-1}$~\cite{Israel,hasan2}.
%%%%%%%%%%%
%\section{Derivation of non-linear equations}
%\label{sec5}
The above hydrodynamic equations from IS theory is used to derive the non-linear equations in (1+1)D~\cite{Fogaca2}. To achieve this we have adopted the Reductive Perturbative Method (RPM)~\cite{RPM1, RPM2, RPM3}, with
``stretched co-ordinates'' defined as, 
\beqa
X=\frac{\sigma^{1/2}}{L}(x-c_s t) \hspace{0.3cm}\text{and} \hspace{0.4cm} Y=\frac{\sigma^{3/2}}{L}(c_s t)
\eeqa
where, $L$ is the characteristic length, $c_{s}$ is 
the speed of sound and $\sigma$ is the expansion parameter. 
Therefore, we have
\beqa
\frac{\partial}{\partial x}=\frac{\sigma^{1/2}}{L}\frac{\partial}{\partial X} \hspace{0.2cm}\text{and} \hspace{0.4cm} 
\frac{\partial}{\partial t}=-c_{s}\frac{\sigma^{1/2}}{L}\frac{\partial}{\partial X}+c_{s}\frac{\sigma^{3/2}}{L}\frac{\partial}{\partial Y}
\eeqa
 The coordinate $X$ is measured from the frame of propagating sound waves, whereas, the $Y$ represents a fast-moving coordinate. 
The RPM technique is devised to preserve the structural form of the parent equation in different order of $\sigma$ {\it i.e} Breaking wave, Burger's, Korteweg–De Vries(KdV) equation {\it etc}. 
Now we do the simultaneous series expansion of hydrodynamic quantities in powers of $\sigma$. For $\epsilon$,
the series appears as:
\beqa
\hat{\epsilon}&=&\frac{\epsilon}{\epsilon_0}=1+\sigma \epsilon_1+\sigma^2 \epsilon_2+\sigma^3 \epsilon_3+...
\label{epshat}
\eeqa
Here, we keep terms up to $\sigma^3$.
We derive the required equations by collecting the 
terms corresponding to the  different orders of 
$\sigma$.  
Finally we revert from $(X,Y) \to (t,x)$ to get the following 
equations for the perturbation in $\hat{\epsilon}$ as:
\begin{widetext}
\beqa
\frac{\pd \hat{\epsilon_{1}}}{\pd t}+[1+(1-c_{s}^2) \frac{\epsilon_{0}}{\epsilon_{0}+p_{0}}\hat{\epsilon_{1}}]c_{s}\frac{\pd \hat{\epsilon_{1}}}{\pd x}- [\frac{1}{2(\epsilon_{0}+p_{0})}(\zeta +\frac{4}{3}\eta)]\frac{\pd^{2} \hat{\epsilon_{1}}}{\pd x^{2}}=0,
\label{nlw1}
\eeqa
%%%%
and
\beqa
\frac{\pd \hat{\epsilon_{2}}}{\pd t}+\mathcal{S}_{1}\frac{\pd \hat{\epsilon_{2}}}{\pd x}+\mathcal{S}_{2}\frac{\pd \hat{\epsilon_{1}}}{\pd x}+\mathcal{S}_{3} \frac{\pd^{2} \hat{\epsilon_{1}}}{\pd x^{2}}+\mathcal{S}_{4}\frac{\pd^{3} \hat{\epsilon_{1}}}{\pd x^{3}}+\mathcal{S}_{5}\frac{\pd^{2} \hat{\epsilon_{2}}}{\pd x^{2}}=0
\label{nlw2}
\eeqa
where the coefficients, $\mathcal{S}_i$'s for $i=1$ to $5$ are given by,
\beqa
\mathcal{S}_{1}&=&\frac{1}{\epsilon_{0}+p_{0}}  \Big[c_s \{\epsilon _0 \left(1- \hat{\epsilon }_1\left(c_s^2-1\right)\right)+p_0\}\Big];\,\, \mathcal{S}_{2}= \frac{1}{\epsilon_{0}+p_{0}}  \Big[\epsilon _0 c_s \{c_s^2-1\} \{\epsilon _0 \left(\left(2 c_s^2+1\right) \hat{\epsilon }_1{}^2-\hat{\epsilon }_2\right)-p_0 \hat{\epsilon }_2\}\Big];\nn\\
\mathcal{S}_{3}&=& \frac{1}{12(\epsilon_{0}+p_{0})^{2}}  \Big[\epsilon _0 \hat{\epsilon }_1 \{3 c_s^2 (7 \zeta +8 \eta )+3 \zeta +4 \eta \}\Big];\,\, \mathcal{S}_{4}=- \frac{1}{72 c_s c_V (\epsilon _0+p_{0})^2}
\Big[4 c_{V} c_s^2 \{3 \kappa  T \epsilon _0(3 \alpha _0 \zeta +4 \alpha _1 \eta )\nn\\
&&+3\kappa T(3 \zeta +4 \eta) +3 \kappa T p_0 (3 \alpha _0 \zeta +4 \alpha _1 \eta )+(\epsilon _0+p_{0}) (9 \beta _0 \zeta ^2+16 \beta _2 \eta ^2)\}-c_{V}(3 \zeta +4 \eta )^2\nn\\
&&+12 \kappa  \{\epsilon _0+p_{0}\} \{\epsilon _0 (3 \alpha _0 \zeta +4 \alpha _1 \eta )+3 \zeta +4 \eta +p_0 (3 \alpha _0 \zeta +4 \alpha _1 \eta )\}\Big];\,\,  \mathcal{S}_{5}  =-\frac{3 \zeta +4 \eta }{6 \left(p_0+\epsilon _0\right)}
\eeqa
\end{widetext}
with, $\hat{\epsilon_{1}}= \sigma \epsilon_{1}$ and 
$\hat{\epsilon_{2}}=\sigma^{2} \epsilon_{2}$.
We observe that 
Eq.~\ref{nlw1} does not contain any effect of the second-order 
theory (IS) as earlier observed in Refs.\cite{Fogaca1, Fogaca2} 
with conformal EoS. This equation can be derived from the NS theory as well. However, the second equation contains the second-order effects via relaxation and coupling coefficients of IS theory. 
We take into account all the transport coefficients 
to provide a general equation for the propagation of 
the nonlinear waves. 
The dispersive terms in Eq.\eqref{nlw2} indicate that the 
combined effects 
of shear and bulk viscosities act 
against the effect of thermal conductivity. 
This might dilute the diffusion of nonlinear waves in a dissimilar way.\\\\
%%%%%%%%%
%%%%%%%%%
%\section{Results and Discussions}
%\label{sec6}
{\it \textbf{III. Results and Discussions}: } In this section we discuss 
the effects of the CEP on the propagation of nonlinear waves. 
We assume that the CEP is located at $(T_{c}, \mu_{c}) = (154, 367)$ MeV, where $\mu_c$ and $T_c$ are the critical values of baryonic chemical potential and temperature respectively. The initial profile of the perturbations is taken as:
\begin{equation}
\hat{\epsilon_{i}}=A_{i} \Big[sech(\frac{x-10}{B_{i}})\Big]^{2}
\label{initialperurb}
\end{equation}
where $i= 1$ and $2$ stand for first and second order perturbations 
respectively with $A_{i}$ and $B_{i}$ 
determining the height and width of the initial profile. 
%In the present work we take  $A_{1}=0.6, B_{1}=1, A_{2}=0.4$ and $B_{2}=1$.
The  effects of the CEP have been taken into consideration through 
the EoS and the scaling behaviour of  transport coefficients and 
Of the thermodynamic response functions.  
The CEP in QGP-hadron transition belongs to the same universality class as that of the  
3D Ising model. 
The construction of EoS with the CEP have been studied in 
Refs.\cite{Asakawa2,parotto,mitedu,Guida} (for details we 
refer to these references). The  procedure  
discussed  in Refs.\cite{Asakawa2,parotto} has been 
followed to construct the EoS with the CEP and used in
 Refs.~\cite{hasan1,hasan2} to study linear perturbation. 
The same EoS has been used in the present work for 
the evolution of nonlinear perturbation. 
%{\color{red} Though there have been progress in constructing the EoS 
%to include the CEP, we use this EoS to capture the main aspect of the 
%CEP e.g. enhancement of transport coefficients, droping speed of sound. 
%It is seen in the linear analysis~\cite{hasan1,hasan2} that effect of 
%EoS is much more decisive in affecting the propagation of perturbation 
%near the CEP than the diverging transport coefficient. For the same purpose,}
The critical behaviour of various  transport coefficients and response
functions are taken from Refs.~\cite{kunihiro,Guida,Kapusta}.

%%%%%%%%%%%%%%%%%%%%%%%%%
%\begin{widetext}
\begin{figure*}
%\centering
\includegraphics[width=5.5cm]{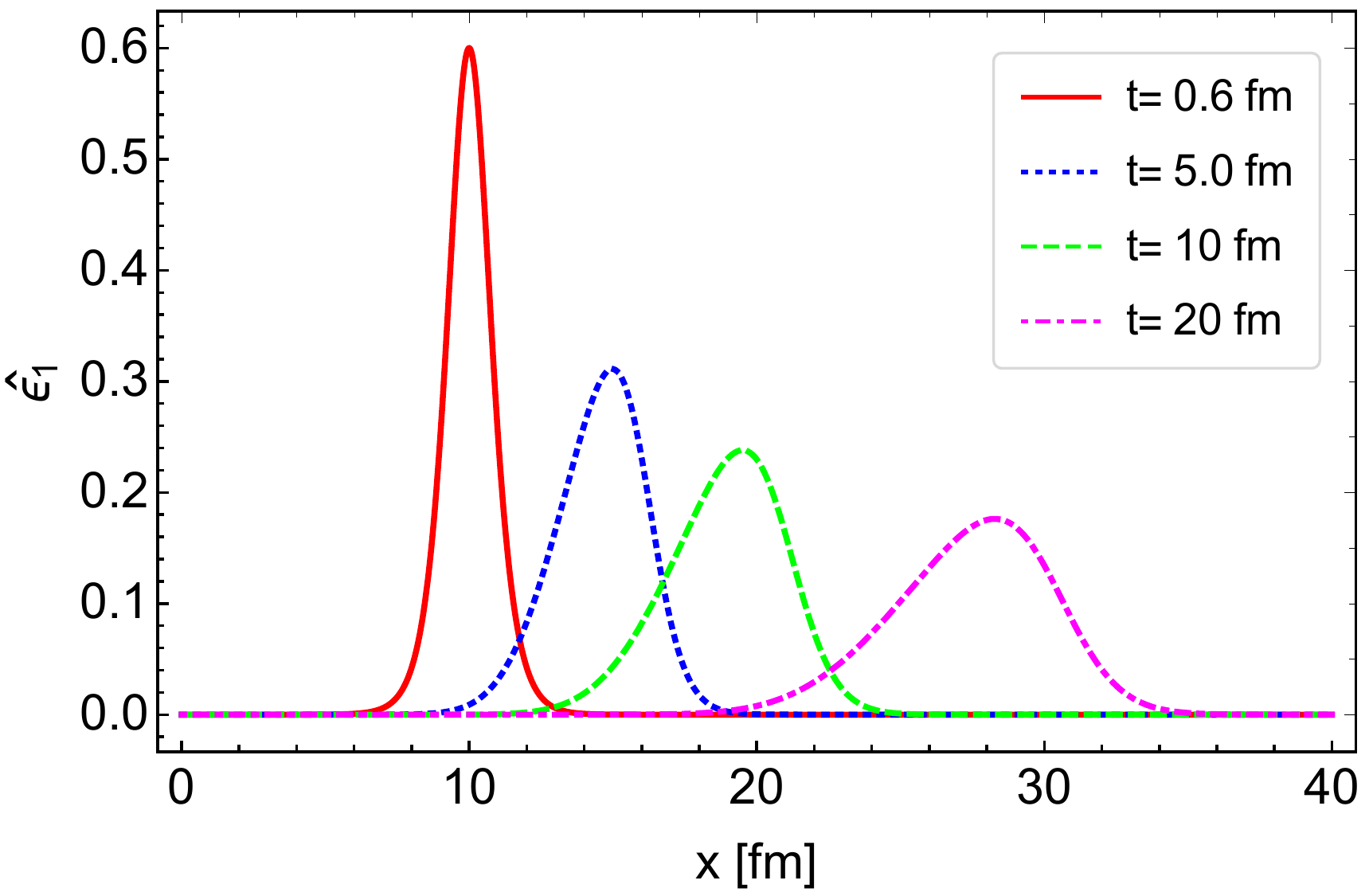}
\includegraphics[width=5.5cm]{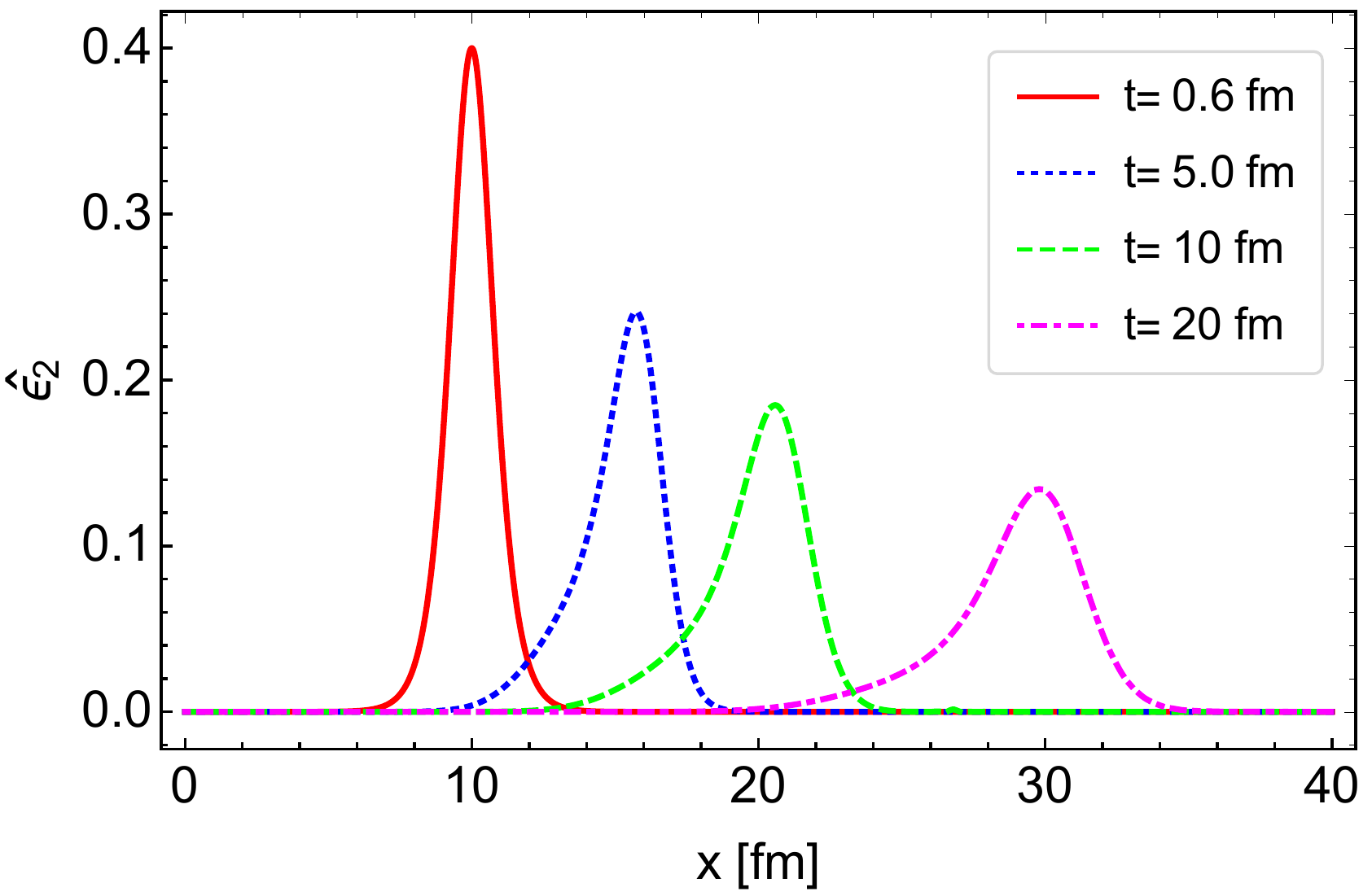}
\includegraphics[width=5.5cm]{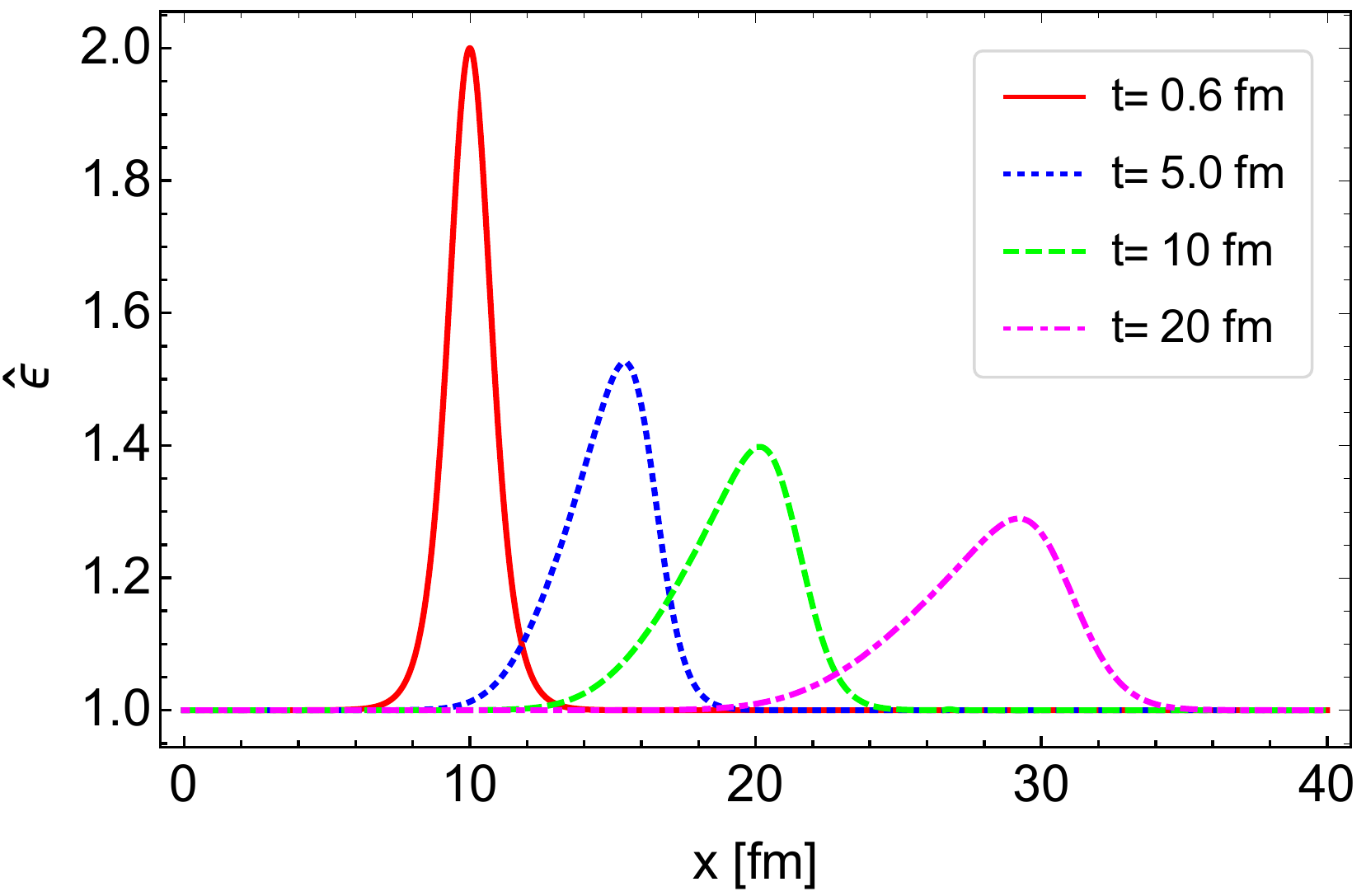}
\caption{(Color online) The spatial variation of perturbations
in energy density when the system is away 
from the CEP at $\mu=400$ MeV and $T=185$ MeV.
The left panel shows $x$ dependence of $\hat{\epsilon_1}=
\sigma\epsilon_1$.
The middle and right panels indicate the variation 
of $\hat{\epsilon_2}=\sigma^2\epsilon_2$ and $[1+\hat{\epsilon_1}+
\hat{\epsilon_2}]$ (see Eq.~\ref{epshat}) respectively with $x$.
It is clear that when the system is 
away from the CEP, the perturbations survive despite the dissipative 
effects. 
The results are obtained here for 
$A_{1}=0.6, B_{1}=1, A_{2}=0.4$ and $B_{2}=1$.
}
\label{fig1}
\end{figure*}
%\end{widetext}
%%%%%%%%%%%%%%%%
	%%%%%%%%%%%%%%%%
%\begin{widetext}
\begin{figure*}
%\centering
\includegraphics[width=5.5cm]{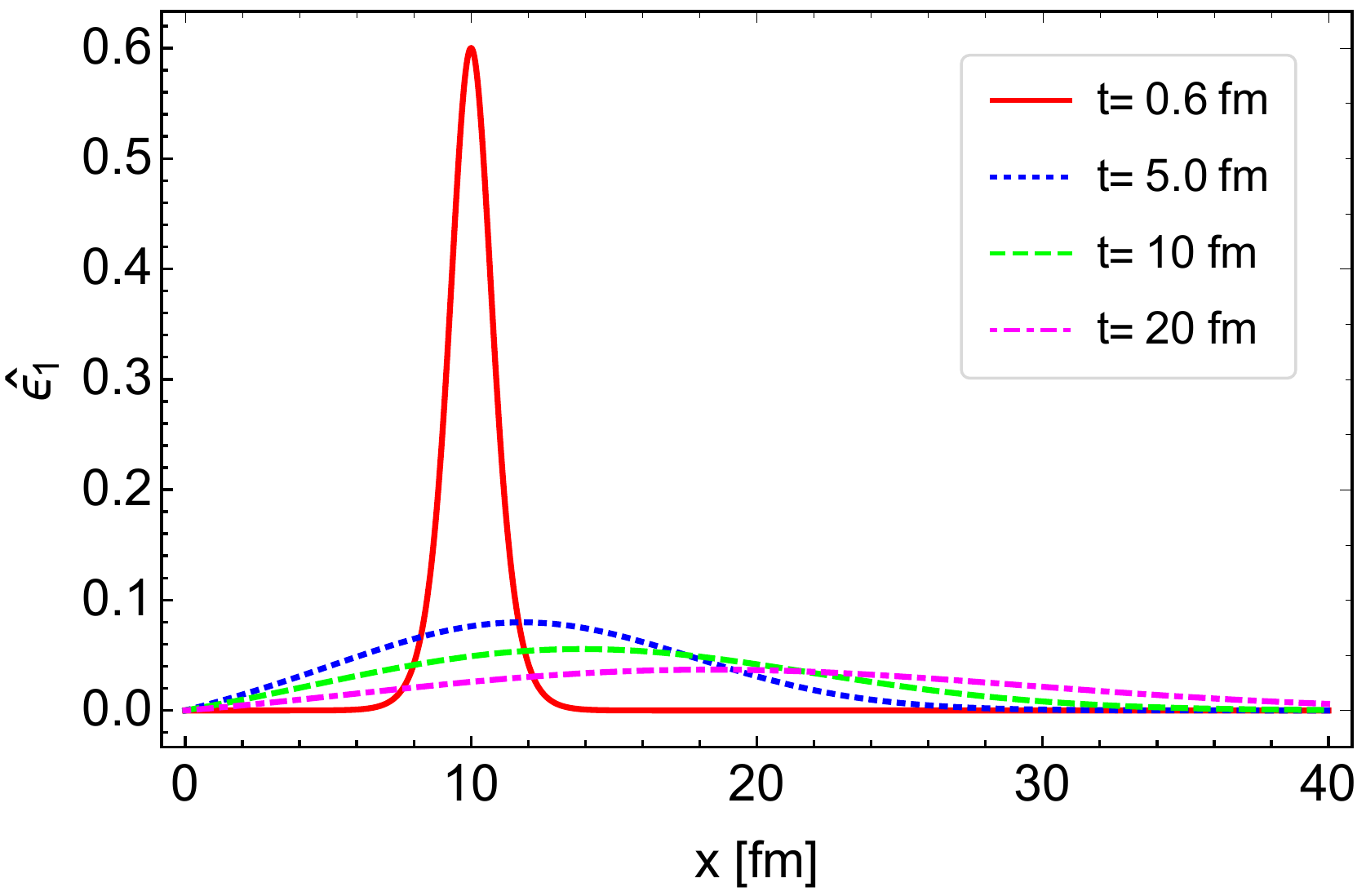}
\includegraphics[width=5.5cm]{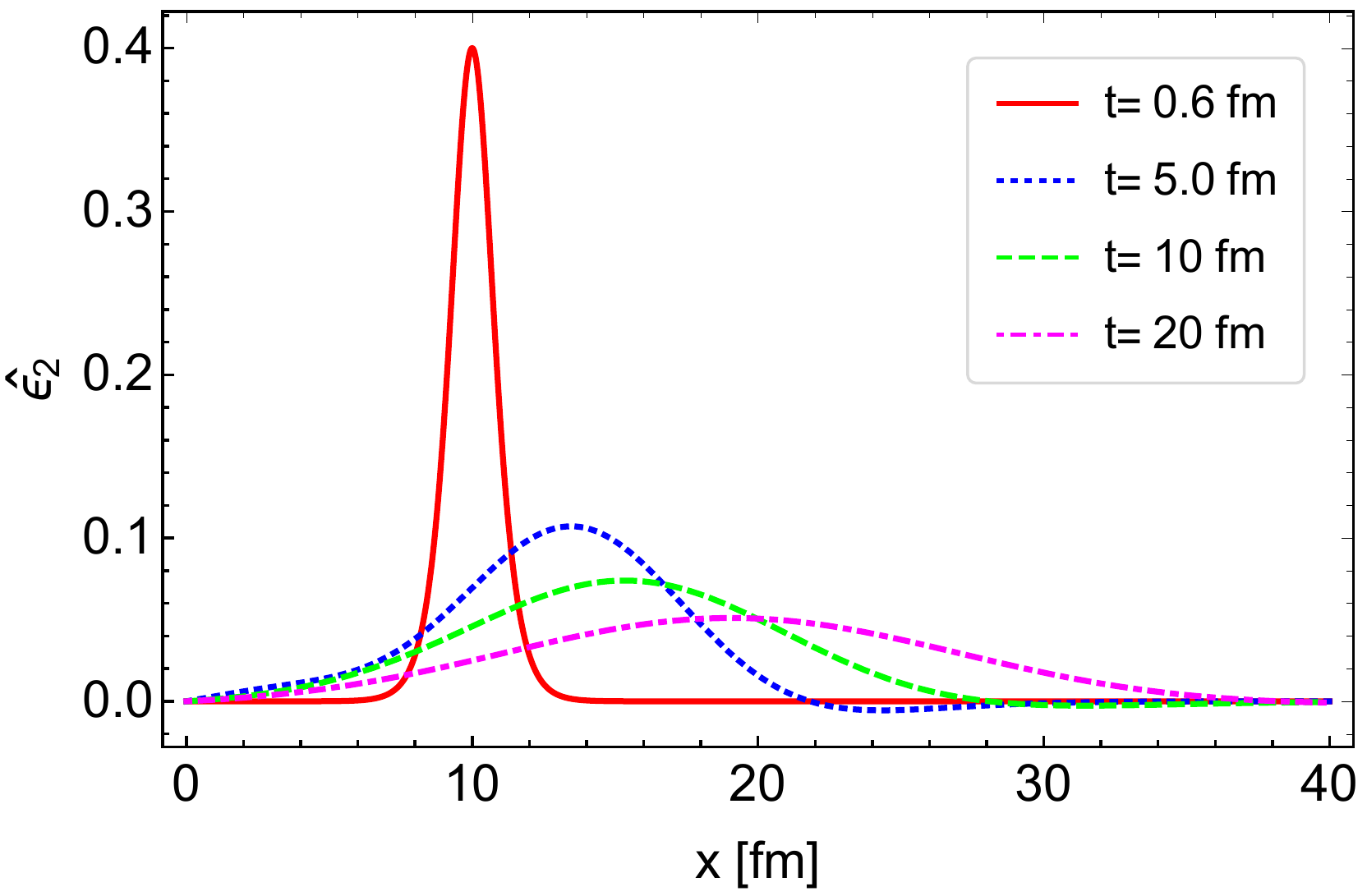}
\includegraphics[width=5.5cm]{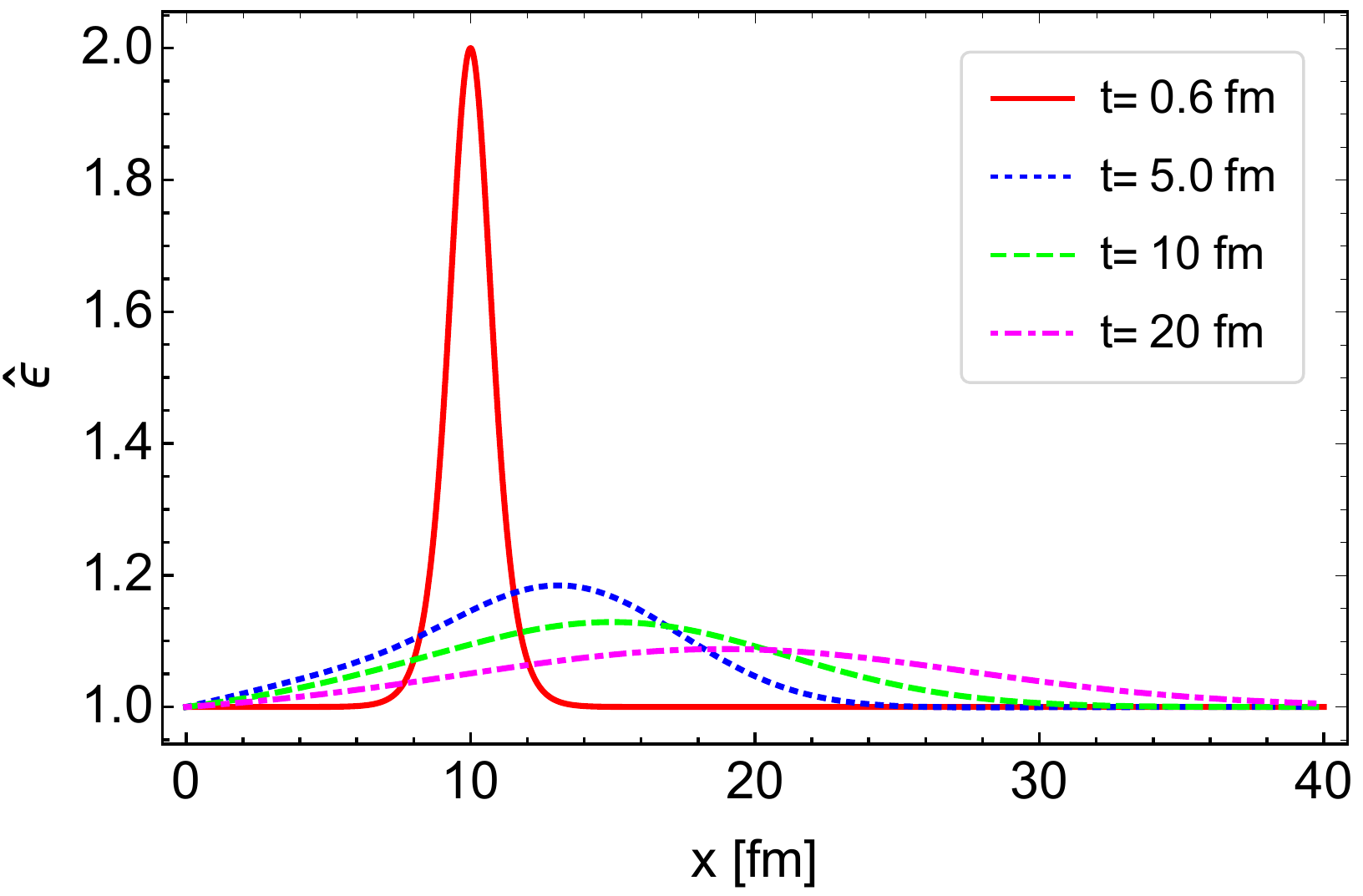}
\caption{(Color online) The spatial variation of perturbations
in energy density when the system is close to the CEP at $\mu=367$ MeV and $T=158$ MeV.
The left panel shows $x$ dependence of $\hat{\epsilon_1}=
\sigma\epsilon_1$.
The middle and right panels indicate the variation 
of $\hat{\epsilon_2}=\sigma^2\epsilon_2$ and $[1+\hat{\epsilon_1}+
\hat{\epsilon_2}]$ (see Eq.~\ref{epshat}) respectively with $x$.
It is clear that when the system is 
near the CEP, the nonlinear perturbations is strongly dissipated. 
The results are obtained here for 
$A_{1}=0.6, B_{1}=1, A_{2}=0.4$ and $B_{2}=1$.
}
\label{fig2}
\end{figure*}
%\end{widetext}
%%%%%%%%%%%%%%%%

Fig.\ref{fig1} shows the propagation of a nonlinear wave through 
the medium when 
the system is formed away from the CEP. It is found that
the nonlinear wave survives with reduced amplitude  
despite the presence of dissipative effects introduced 
via the non-zero values of transport coefficients.  
In contrast, the nonlinear 
wave is substantially dissipated near the CEP 
as evident from the results depicted in Fig.~\ref{fig2}. 
It has been observed that the
the dissipative feature remains unchanged with the variation of the position of the CEP along 
the transition line in the QCD phase diagram.
This clearly indicates that the nonlinear perturbations will provide
detectable effects of the CEP.
It is to be noted that the speed of the 
nonlinear wave is non-zero
in contrast to linear waves~\cite{hasan1}. This is due to the amplitude-dependent propagation speed of the nonlinear waves. 
The attenuation of the perturbation is smaller for
the second-order correction ($\hat{\epsilon_2}$).
Furthermore, the second-order perturbation travels a bit faster tan the linear one. 
This is clear from the results displayed in Figs.\ref{fig1} and Fig.\ref{fig2} 
which can be understood from Eq.\eqref{nlw2}, where the second-order correction 
contains the third-order derivatives of first-order perturbation.
This is similar to the dispersive term in the KdV equation responsible for 
height preserving solitonic behaviour\cite{Lick1970}. 
Therefore, dispersive terms compete with the diffusive terms 
to weakening the damping effect. 
The relaxation effects on the dissipative fluxes incorporated in IS theory
(Eq.\eqref{nlw2}) makes the dissipation slower in 
comparison to the NS theory (Eq.\eqref{nlw1}).  
It is also found that 
the diverging nature of thermal conductivity near the CEP~\cite{Kapusta} 
dominates over the shear and bulk viscous effects as the
the degree of divergence of $\kappa$ is stronger than 
the other transport coefficients.
%This is similar to the case of linear perturbations~\cite{hasan1}. 
%The contribution from $\kappa$ gets more weight due to presence of $1/c_{V}$ 
%was factor to $\kappa$ in the dispersive term with fractional value of $c_{V}$. 

Therefore, near the CEP the waves with large amplitudes created in the medium by the energetic particles or jets, will be highly suppressed. It has been seen earlier that the 
linear waves are fully halted near the CEP. 
We find similar suppression of nonlinear waves due to the presence of the CEP.
Although the propagation speed of nonlinear waves are amplitude dependent,
they are highly suppressed near the CEP irrespective of their amplitude.  
This has important consequences in detecting the CEP.
 
It has been predicted earlier~\cite{kunihiro,hasan1,hasan2} that the formation 
of Mach cone is prevented near the CEP for linear perturbation. 
Whether a similar effect is observed for the nonlinear 
waves, is an interesting question to address. This is crucial because nonlinear effects are found to be shape and height-preserving in comparison to linear propagation. 
In this work, we find that even the nonlinear perturbations will not be able to retain the Mach cone effects if it hits the CEP. 
%So the propagation of both the perturbations 
%(linear and nonlinear) is stopped due to the presence of CEP.  
These findings can be used to detect the CEP by looking into the suppression of Mach cone in two-particle correlation~\cite{Betz1}.
Due to the suppression of nonlinear waves, the broadening 
effect of localized waves will also vanish ~\cite{Fogaca1}
along with the Mach cone effect.
%%%%%%%%%%%%%%%%%%%%%%%%%%%%%%%%%%%%%%%%%%%%%%%%%%%%%%%%%%%%%%%%%%%%%%%%%%%%%%%%

The flow harmonics play a crucial role in characterizing the medium formed in 
RHIC-E. It was shown in Refs. ~\cite{stocker1, stocker2}
that some of the flow harmonics will collapse at the CEP. 
The present work has crucial consequences on the flow harmonics. Based on linear analysis it has been argued in
Refs.~\cite{kunihiro,hasan1,hasan2}  
that $v_{2}$ will be reduced near the CEP. The same conclusion can be drawn for the nonlinear waves too in presence of the CEP. 
If the initial spatial shape of the system formed in HICs is highly distorted 
azimuthally, which may be the case for off-central collisions,  
the fluid dynamical response to the initial eccentricity may be nonlinear. 
The present work suggests that $v_2$ and higher harmonics will be suppressed near 
the CEP due to the absorption of the sound wave, even for highly off-central collisions. %Since the internal 
%inhomogeneity is higher in magnitude, nonlinear effects may be triggered, and 
%will also be suppressed in the vicinity of CEP.

In a recent work~\cite{Dore}, it is found that the path to the critical point is heavily 
influenced by far from equilibrium initial conditions, 
where viscous effects lead to dramatically different $(T,\mu)$ 
trajectories. 
This means that the trajectory of the system in the QCD phase diagram will be event-dependent. Therefore, it will be useful
to analyze the flow harmonics on event-by-event basis. If a particular event evolve through the CEP, then the rms (root mean square)
value of flow harmonics will be suppressed because of the
absorption of the sound wave. On the contrary,
if an event evolves through a trajectory that is away from the
CEP, then the flow harmonics will survive. Hence, 
the presence of the CEP will cause large event-by-event
fluctuations of flow harmonics.

We emphasize that the nonlinear perturbations damped substantially if the system passes through the CEP. In absence of the critical point the nonlinear perturbation survives although the medium is dissipative with non-zero $\eta, \zeta$ and $\chi$. This indicates that the formation of Mach cone will be prohibited in the presence of the CEP. Therefore, the conclusion of the present theoretical investigation is that the Mach cones disappear in the presence of the CEP. The effects of Mach cone manifest as a double hump in the two-particle correlation in the low momentum domain of associated particles {\it i.e.} the CEP plays a unique role in suppressing the double hump in the two-particle correlation contrary to the other mechanisms: {\it e.g.} (i) deflection of away side jets, (ii) Cherenkov radiation and (iii) radiation of gluons which produce the double hump. These mechanisms have the ability to obscure the suppression due to the CEP by creating double hump.

%%%%%%%%%%
The observation~\cite{mach1,mach2} of the dip in the azimuthal distribution of
two particle correlation at $\Delta\phi (=\phi-\phi_{trig})=\pi$
accompanied by two local peaks on either side of $\Delta\phi=\pi$  
for transverse momentum range, $0.15< p_T^{assoc} < 4$ GeV,
is attributed to the physical processes mentioned above.
Therefore, we contrast the effect of the CEP to the following
mechanisms.
(i) Deflection of the away side jets by strong 
asymmetric flow in non-central collisions 
and third flow harmonics due to initial state fluctuations 
(Refs.~\cite{wang,Scao,Betz1} for further details) lead to
peaks of the away side jet on the either side of $\Delta\phi=\pi$. 
However, if the system passes through the CEP, the flow will be highly suppressed and hence,
the deflection too will be strongly reduced. (ii)  Cherenkov radiation ~\cite{cerenkov1,cerenkov2}
is characterized by strong momentum
dependence of the cone angle ($\sim 1/\mu$ where $\mu$ is the refractive index of the medium). 
This process is unlikely to be responsible for the double hump because of the
lack of observed momentum dependence of the location of the double peaks of associated particles. 
(iii) The radiation of gluons by the away side jet will deviate it from propagating at an angle
180$^\circ$ with respect to the near side (trigger) jet. 
However, the quantitative prediction of the Mach cone positions studied through
three-particle correlation~\cite{3particle1,3particle2} and the momentum independence of the
location of the double hump 
indicate that the double hump may originate from Mach cone effects. 
The vanishing of the Mach-cone like structure in particle correlation will 
therefore, indicate the existence of the CEP.

In this study, the propagation of perturbation is studied in a static background. In presence of expansion, the system will cool and move towards the phase transition line with changing transport coefficients. However, during this cooling, if the trajectory goes away from the CEP, 
then the perturbations will still survive. 
But if it passes near the CEP, the 
perturbative effects will be mostly washed out and no effect will survive at the latter stage. 
Therefore, it is expected that the results obtained in the static situation may not
vary with the inclusion of expansion if the system passes 
through the CEP. We have investigated the hydrodynamic propagation of perturbations 
in the system without considering the 
fluctuations~\cite{hydro+1,hydro+2} originating from the CEP itself, 
as they do not create any angular pattern as jets produce in correlations.
Although the non-equilibrium fluctuations due to the CEP is not taken into 
the account here for obtaining the non-linear wave equations, the enhancement 
of thermodynamic fluctuations near the critical point which affects 
the hydrodynamic response is inherently taken into account through 
the EoS and other thermodynamic quantities via the critical exponents. \\\\
%\section{Summary and Conclusion}
%\label{sec7}

{\it \textbf{IV. Summary and Conclusion}:} 
In summary, we have investigated the response of the
QCD critical point to the nonlinear perturbations 
within the scope of second-order IS hydrodynamics. The effects
of the CEP on the propagation of nonlinear waves have been taken into accounts
through the EoS, critical behavior of the transport
coefficients and of thermodynamic response functions.  
We have derived relevant equations governing the 
propagation of nonlinear waves within the purview
of second-order causal hydrodynamics 
by taking into account the non-zero values of 
$\eta$, $\zeta$ and $\kappa$
in contrast to earlier works where 
the effects of $\zeta$ and $\kappa$ were ignored. 
In the presence of the CEP $\zeta$ and $\kappa$ play important
roles as they diverge near the CEP and hence can not be ignored.
It is found that, similar to the linear perturbation,
the nonlinear perturbations too get suppressed near the 
CEP. The diverging nature of thermal conductivity near the CEP
plays the most dominant role in the suppression of nonlinear waves. 
The nonlinear effects make the perturbation to  
travel a bit faster than the linear one. 
The presence of the CEP will be resulting in the vanishing of Mach cone effects (or away side double-peak structure) and 
the broadening of the two and three-particle correlation.
The suppression or collapse of elliptic flow will also indicate the
existence of the CEP. This may lead to
the large event-by-event fluctuation of flow harmonics between
two events with and without the CEP.
Therefore, the vanishing Mach cone effects 
(or away side double-peak structure) on the away side jet and 
the enhancement of fluctuation of flow harmonics in event-by-event analysis accompanied by 
suppressed flow harmonics could be considered as signals of the CEP.

 {\it \textbf{Declaration of competing interest:}} 
The authors declare that they have no known competing financial interests or personal relationships 
that could have appeared to influence the work reported in this paper.
%%%%%%%
%%%%%%%
%%%%%%%
%\section{Acknowledgement}
\section{Acknowledgement}
AB thanks Alexander von Humboldt (AvH) foundation and Federal
Ministry of Education and Research (Germany) for support
through Research Group Linkage programme.
 
 % Uncomment the following two lines if you want to have a bibliography


\begin{thebibliography}{alpha}

\bibitem{Forcrand} P.~de Forcrand and O.~Philipsen, Nucl. Phys. B \textbf{642}, 290 (2002).
%``The QCD phase diagram for small densities from imaginary chemical potential,''

\bibitem{Aoki}
Y.~Aoki, G.~Endrodi, Z.~Fodor, S.~Katz and K.~Szabo, Nature \textbf{443}, 675 (2006).
%``The Order of the quantum chromodynamics transition predicted by the standard model of particle physics,''

\bibitem{Endrodi}
G.~Endrodi, Z.~Fodor, S.~Katz and K.~Szabo, JHEP \textbf{04}, 001 (2011).
%%

\bibitem{PNJL10} A. Bhattacharyya, P. Deb, 
S. K. Ghosh and R. Ray, Phys. Rev. D {\bf 82}, 014021 (2010).

\bibitem{PQM1} B.~J.~Schaefer, J.~M.~Pawlowski and J.~Wambach, 
Phys.\ Rev.\ D {\bf 76}, 074023 (2007).


\bibitem{Fodor}
Z.~Fodor and S.~Katz, JHEP \textbf{03}, 014 (2002).
%``Lattice determination of the critical point of QCD at finite T and mu,''

\bibitem{AsakawaYazaki}
M.~Asakawa and K.~Yazaki, Nucl. Phys. A \textbf{504}, 668 (1989).
%``Chiral Restoration at Finite Density and Temperature,''

\bibitem{Halasz}
A.~M.~Halasz, A.~Jackson, R.~Shrock, M.~A.~Stephanov and
J.~Verbaarschot, Phys. Rev. D \textbf{58}, 096007 (1998).

\bibitem{Caines} H. Caines, Nuclear Physics A {\bf 967}, 121 (2017).

\bibitem{NahrangBluhm}
M.~Nahrgang, M.~Bluhm, T.~Schaefer and S.~A.~Bass,
%``Diffusive dynamics of critical fluctuations near the QCD critical point,''
Phys. Rev. D \textbf{99}, 116015 (2019), M.~Bluhm, A.~Kalweit, M.~Nahrgang, M.~Arslandok, P.~Braun-Munzinger, S.~Floerchinger, E.~S.~Fraga, M.~Gazdzicki, C.~Hartnack and C.~Herold, \textit{et al.}
%``Dynamics of critical fluctuations: Theory \textendash{} phenomenology \textendash{} heavy-ion collisions,''
Nucl. Phys. A \textbf{1003}, 122016 (2020).


\bibitem{Stanley} H. E. Stanley, Introduction to phase transitions and critical phenomena, Oxford University Press, 1971. 



%\bibitem{Huang} K. Huang, Statistical Mechanics, John Wiley \& Sons, New York, 1987.

\bibitem{Kardar} M. Kardar, Statistical Physics of Fields, Cambridge University Press, 2007.

\bibitem{Shankar} R. Shankar, Quantum Field Theory and Condensed Matter:  An Introduction, Cambridge University Press, 2017.

\bibitem{Berdnikov} B. Berdnikov and K. Rajagopal, Phys. Rev. D {\bf 61} 105017 (2000).

%\bibitem{Stephanov1} M. Stephanov, K. Rajagopal and E. Shuryak, Phys. Rev. D {\bf 60} 114028 (1999).

\bibitem{Stephanov2} 
M. Stephanov, K. Rajagopal and E. Shuryak, Phys. Rev. Lett. {\bf 81}, 4816 (1998).

\bibitem{shuryaknlw} J. C. Solana, E. Shuryak and D. Teaney, J. Phys.: Conf. Ser. 
{\bf 27}, 003 (2005).

%%%
\bibitem{Shuryak1} E. Shuryak, Phys. Rev. C {\bf 80}, 069902 (2009).

\bibitem{Shuryak2} P. Staig and 
E. Shuryak, Phys. Rev. C {\bf 84}, 034908 (2011), Phys. Rev. C {\bf 84}, 044912 (2011).

\bibitem{Rafiei:2016zxk}
A.~Rafiei, K.~Javidan,
%``Colliding solitary waves in quark gluon plasmas,''
Phys. Rev. C \textbf{94},  034904 (2016).

\bibitem{hasan1} M. Hasanujjaman, M. Rahman, A. Bhattacharyya and 
J. Alam, Phys. Rev. C {\bf 102}, 034910 (2020).

\bibitem{hasan2} M. Hasanujjaman, G. Sarwar, M. Rahman, A. Bhattacharyya and 
J. Alam, arXiv:2008.03931v2.

\bibitem{kunihiro} Y. Minami and T. Kunihiro, Prog. Th. Phys. {\bf 122}, 881 (2009). 


\bibitem{Raha1} S. Raha, K. Wehrberger and 
R.M. Weiner, Nucl. Phys. A {\bf 433}, 427 (1984).

\bibitem{Raha2} G.N. Fowler, S. Raha, N. Stelte and R.M. Weiner, Phys. Lett. B 
{\bf 115}, 286 (1982); S. Raha and R. M. Weiner,
Phys. Rev. Lett. {\bf 50}, 407 (1983); 
E. F. Hefter, S. Raha and R. M. Weiner, 
Phys. Rev. C {\bf 32}, 2201 (1985).

\bibitem{Fogaca1} D. A. Fogaca, L. G. Ferreira Filho
and F. S. Navarra, Phys. Rev. C {\bf 81}, 055211 (2010).

\bibitem{Fogaca2}
D.A. Fogaca, H. Marrochio, F.S. Navarra and J. Noronha, Nucl. Phys. A {\bf 934}, 18 (2015).

%\bibitem{Trambak} T. Bhattacharyya and A. Mukherjee, Eur. Phys. J. C {\bf 80}, 656 (2020).

%%%%

\bibitem{v2e2} H. Niemi, G. S. Denicol, H. Holopainen and P. Huovinen,
Phys. Rev. C{\bf 87}, 054901 (2013).

\bibitem{v2ne2} G. Giacalone, J. Noronha-Hostler and J. Y. Ollitrault, 
Phys. Rev. C {\bf 95}, 054910 (2017).

\bibitem{v2ne22} The ALICE collaboration., S. Acharya et al., 
JHEP {\bf 2020}, 85 (2020).
%%%%%%

\bibitem{akc} A. K. Chaudhuri and U. Heinz, Phys. Rev. Lett. {\bf 97}, 062301 (2006).

\bibitem{Osbourne} M. F. M. Osborne and A. H. Taylor, Phys. Rev. {\bf 70}, 322 (1946).

\bibitem{Betz1} B. Betz, arXiv:0910.4114.

\bibitem{Betz2} B. Betz, J. Noronha, G. Torrieri, M. Gyulassy and D. H. Rischke, 
Phys. Rev. Lett. {\bf 105}, 222301 (2010).


\bibitem{Li} H. Li, F. Liu, G. Ma, X. N Wang and Y. Zhu, Phys. Rev. Lett. 
{\bf 106}, 012301 (2011).

\bibitem{Israel} W. Israel and J. M. Stewart, Ann. Phys. (NY) {\bf 118}, 341 (1979).

\bibitem{Kharzeev} D. Kharzeev and K. Tuchin, JHEP {\bf 09}, 093 (2008).

\bibitem{Karsch}
F. Karsch, D. Kharzeev and K. Tuchin, Phys. Lett. B {\bf 663}, 217 (2008).

\bibitem{Ryu} S. Ryu, J. F. Paquet, C. Shen, G. S. Denicol, B. Schenke, 
S. Jeon and C. Gale, Phys. Rev. Lett. {\bf 115}, 132301 (2015).

\bibitem{Kapusta} J. I. Kapusta and J. M. T. Rincon, Phys. Rev. C {\bf 86}, 054911 (2012).

\bibitem{Martinez}
M.~Martinez, T.~Sch\"afer and V.~Skokov,
%``Critical behavior of the bulk viscosity in QCD,''
Phys. Rev. D \textbf{100}, 074017 (2019).

\bibitem{Bemfica}
F.~S.~Bemfica, M.~M.~Disconzi and J.~Noronha, Phys. Rev. D \textbf{98}, 104064 (2018), Phys. Rev. D \textbf{100}, 104020 (2019).

\bibitem{Kovtun}
P.~Kovtun,
%``First-order relativistic hydrodynamics is stable,''
JHEP \textbf{10}, 034 (2019).

\bibitem{Das}
A.~Das, W.~Florkowski, J.~Noronha and R.~Ryblewski,
%``Equivalence between first-order causal and stable hydrodynamics and Israel-Stewart theory for boost-invariant systems with a constant relaxation time,''
Phys. Lett. B \textbf{806}, 135525 (2020).

\bibitem{arpan} A. Das, W. Florkowski and 
R. Ryblewski, Phys. Rev. D {\bf102}, 031501 (2020), Phys. Rev. D {\bf103}, 014011 (2021).

\bibitem{Landau} L. D. Landau and E. M. Lifshitz, Fluid Mechanics (Addison-
Wesley, Boston, 1959).

\bibitem{Eckart} C. Eckart, Phys. Rev. {\bf 58}, 919 (1940).


\bibitem{muronga} A. Muronga, Phys. Rev. C {\bf 69}, 03490 (2004);
Phys. Rev. Lett. {\bf 88}, 062302 (2002); Phys. Rev. Lett. {\bf 89},
159901 (2002) (erratum).

\bibitem{RPM1} H. Washimi and T. Taniuti, Phys. Rev. Lett. {\bf 17}, 996 (1966).

\bibitem{RPM2} R.C. Davidson, “Methods in Nonlinear Plasma Theory”, Academic Press, New York an London, (1972).

\bibitem{RPM3} H. Leblond, J. Phys. B: At. Mol. Opt. Phys. {\bf 41}, 043001 (2008).

%\bibitem{sarwar} G. Sarwar, M. Hasanujjaman, M. Rahman, 
%A. Bhattacharyya, J. Alam, arXiv:2012.12668v2.

\bibitem{Asakawa2} C. Nonaka and M. Asakawa, Phys. Rev. C {\bf 71}, 044904 (2005).

\bibitem{parotto} P. Parotto, M. Bluhm, D. Mroczek, M. Nahrgang,
J. Noronha-Hostler, K. Rajagopal, C. Ratti, T. Schafer and M. Stephanov,
Phys. Rev. C {\bf 101}, 034901 (2020).

\bibitem{mitedu} W. Assawasunthonnet,\\ https://dspace.mit.edu/handle/1721.1/51611.

\bibitem{Guida} R. Guida and J. Zinn-Justin, Nuclear Physics B {\bf 489},  626 (1997).


 \bibitem{Lick1970} W. Lick, Nonlinear Wave Propagation in Fluids, Annual Review of Fluid Mechanics 1970 2:1, 113.

\bibitem{stocker1} H.~Stöcker, %``Collapse of flow: Probing the order of the phase transition,''
PoS \textbf{CPOD07}, 025 (2007), doi:10.22323/1.047.0025,
[arXiv:0710.5089 [hep-ph]].

\bibitem{stocker2}  J. Hofmann, H. Stöcker, U. W. Heinz, W. Scheid 
and W. Greiner, Phys. Rev. Lett. {\bf 36}, 88 (1976).

\bibitem{Dore} T. Dore, J. Noronha-Hostler and 
E. McLaughlin, Phys. Rev. D {\bf 102}, 074017 (2020).

%\bibitem{Betz} B. Betz, arXiv:0910.4114 [nucl-ph]

\bibitem{mach1} F. Wang [STAR Collaboration],  Nucl. Phys. A {\bf 774}, 129 (2006). 

\bibitem{mach2} J. Adams et al. [STAR Collaboration], Nucl. Phys. A {\bf 757}, 102 (2005).

\bibitem{wang} F. Wang, Prog. Part. Nucl. Phys. {\bf 74}, 35 (2014).

\bibitem{Scao} S. Cao and X. N. Wang, Rept. Prog. Phys. {\bf 84}, 024301 (2021).

\bibitem{cerenkov1} V. Koch, A. Majumder and X. N. Wang, 
Phys. Rev. Lett. {\bf 96}, 172302 (2006).

\bibitem{cerenkov2} C. Adler et al. [STAR Collaboration], 
 Phys. Rev. Lett. {\bf 90}, 082302 (2003).

\bibitem{3particle1} B. I. Abelev et al. [STAR Collaboration], 
 Phys. Rev. Lett. {\bf 102}, 052302 (2009).

\bibitem{3particle2} J. G. Ulery [STAR Collaboration], 
Int. J. Mod. Phys. E {\bf 16}, 2005 (2007).

\bibitem{hydro+1} M. Stephanov and Y. Yin, Phys. Rev. D {\bf 98}, 036006 (2018).

\bibitem{hydro+2} X. An, G. Basar, M. Stephanov and H-U Yee, Phys. Rev. C 
{\bf 102}, 034901 (2020).

\end{thebibliography}
\end{document}